# Magnetic configuration sweep control in Heliac type stellarators

J.A. Romero, L. Pacios, A. de la Peña, F. Lapayese, E. Ascasíbar

## Abstract

A novel magnetic configuration sweep control system for the TJ-II Heliac device is described.  The system is prepared to establish a reference for plasma, torus or vessel current while the coil configuration currents are swept during a single plasma discharge. It can also be run in a special simulation mode, which does not require powering the TJ-II device, intended for commissioning and experiment preparation purposes. Preliminary tests with the system have shown its ability to perform rotational transform scans while establishing a waveform for torus current independently, despite variable plasma conditions. This opens up new experimental possibilities to study the influence of rotational transform and shear independently.

## I. INTRODUCTION

Stellarators [1]  are high vacuum toroidal devices relying on magnetic fields to confine high temperature plasmas for the purpose of generating thermonuclear fusion energy [2]. In the stellarator concept, a set of external coils produce a vacuum magnetic field with a toroidal helical structure with a particular type of symmetry known as stellarator symmetry [3]. In cylindrical coordinates, and with respect to the symmetry plane $\phi=0$, an stellarator symmetric magnetic field B verifies

$$\begin{pmatrix} B_R & B_\phi & B_Z \end{pmatrix}_{(R,\phi,Z)} = \begin{pmatrix} -B_R & B_\phi & B_Z \end{pmatrix}_{(R,-\phi,-Z)} \qquad (1)$$

The plasma confining region is made up of magnetic field lines with a helical structure that close on themselves after a finite or infinite number of turns around the cylindrical z axis, which is termed the machine axis. The magnetic field line that closes onto itself after exactly one toroidal turn is termed the magnetic axis. The rotational transform ι is defined as the number of turns of a magnetic field line around the magnetic axis per toroidal turn around the machine axis. A common normalisation is the rotational transform per radian, which will be the standard used all through this paper.  The rotational transform has a substantial impact on confinement and stability at several levels, particularly when rotational transform has rational values.  Magnetic field lines always belong to the so-called magnetic surfaces. The plasma confinement region is made of an infinite collection of magnetic surfaces. Some of them are nested, and some of them intersect with themselves. The self- intersecting magnetic surfaces lead to the so-called magnetic islands, which are associated with rational values of the rotational transform.
The external coils are not the only sources of magnetic field. Stellarators may also have net toroidal current arising from the so-called bootstrap effect. The bootstrap current is produced in the presence of a plasma pressure gradient, and is associated with the existence of trapped charged particles in toroidal magnetic confining systems

[4]. There are also internal currents flowing in the plasma arising from many different plasma magneto-hydro-dynamics and turbulent phenomena, which do not necessarily result in net toroidal current. The resulting magnetic field structure, including self-generated plasma magnetic fields, do not always fulfil the stellarator symmetry of the vacuum fields used to confine the plasma, since this symmetry is imposed by the design of the external magnetic field coils generating the configuration and does not take into account the plasma sources. This is particularly true at locations where the magnetic field lines are associated with rational values of the rotational transform [5], where large Magneto Hydro dynamic (MHD) activity may develop [6].

To increase the plasma temperature close to thermonuclear conditions, stellarators require several MW of plasma heating sources [7] . Some of these systems can also drive certain amount of plasma current in the plasma by means on non-inductive processes [9], which in turn affect the final magnetic field structure of the plasma. Power exhaust from the plasma is handled by redirecting the charged particles to external refrigerated plates using a specially shaped magnetic field structure called divertor [8], whose topology is linked to rational values of the rotational transform. Since the radial profile shape of the rotational transform depends directly on the shape of the plasma current profile [10], a plasma current control mechanism is a minimum requirement to ensure the magnetic integrity of the divertor [11], [12]. Some stellarators, such as the TJ-II device under study, are also equipped with a transformer primary circuit coil system, where the plasma acts as the secondary circuit of the transformer. The transformer coils thus provide a mechanism to regulate the net toroidal current circulating in the plasma and modify the rotational transform radial profile.

The TJ-II is a four period flexible heliac type stellarator [13], [14] optimally suited for the study the influence of rotational transform profiles on MHD and confinement. Twisting of the magnetic axis around a central coil system is the main mechanism to achieve the rotational transform in TJ-II, achieving record values as high as 2.4 in short pulses. It is also equipped with a small transformer to induce small amounts of current in the plasma (typically a few kA). This low level of current does not modify the rotational transform value at the plasma boundary significantly, but it does modify the rotational transform profile significantly as we move towards the magnetic axis [15],[16], making it a valuable tool to modify the radial variation of the rotational transform (shear parameter) across the plasma. The TJ-II device has a nominal plasma volume of about 1m$^3$, limited by the intersection of the outermost magnetic filed lines with the vacuum chamber. The nominal magnetic field intensity is 1 Tesla, and the plasma has a bean shaped cross-section with an average minor radius of about 20cm.

The coil arrangement and geometry of the TJ-II device is described in [17], and is briefly summarized here on account of its relevance for the work presented. TJ-II has seven coil systems. Five of the coil systems are essential to shape the vacuum magnetic configuration, namely the TF (toroidal) , VF (vertical,) CC (central) and two Helical conductors, named HX1 and HX2, which wrap around the central ring conductor of 1.5 m radius four times per toroidal turn. With these coil systems, a bean shaped plasma cross-section with a helical magnetic axis that wraps helically around the central conductor is achieved. An additional R (radial) system is available to correct stellarator symmetry breakings arising from coil construction and installation tolerances. The R system is not required for operation, due to the good magnetic surface quality of TJ-II

[18],[19]. The seventh system is the ohmic heating (OH) coil actuator. The OH actuator is the primary of a transformer where the plasma is the secondary. It has a flux swing capability of 0.2 Wb capable of inducing a few kA of ohmic current in the plasma. As advanced earlier, this actuator is used to force the plasma current to follow a certain waveform despite non-inductive currents or inductive voltage drive in the plasma, which is the subject of this work.

The figure 1 shows the coil geometry and resulting plasma wrapping around the central conductor.

All coil systems are made of copper and powered independently by seven 12-pulse thyristor converters with local decoupled control of the coil currents [20] . A flywheel generator provides about 80MW of power with a maximum pulse length of 3 seconds [21]. The plasma discharge takes place on the flat top of the coil configurations currents, and lasts for just few hundred milliseconds, being limited by the maximum temperature that can safely withstand the coil's epoxy insulation without cracking. Plasma discharges can be repeated once every 5–10min, depending on the coil cooling times.

Plasma heating during the flat top is provided by two gyrotrons of 300kW power each, working at the second harmonic electron resonance frequency of 53.2. GHz [22]. The magnitude of the magnetic field $|\vec{B}|$ required for electron cyclotron resonance heating (ECRH) at this frequency is 0.96 T, requiring about 28kA circulating in the toroidal circuit. The same ECH system can be used to drive a few kA of highly localized non-inductive Electron Cyclotron Current drive (ECCD) in the plasma by changing the launching direction of the ECRH beams [23]. In addition TJ-II has two neutral beam injectors (NBI) with 800kW nominal power each [24], and a Bernstein system working at 28Ghz with a nominal power of 200kW [25],[26].

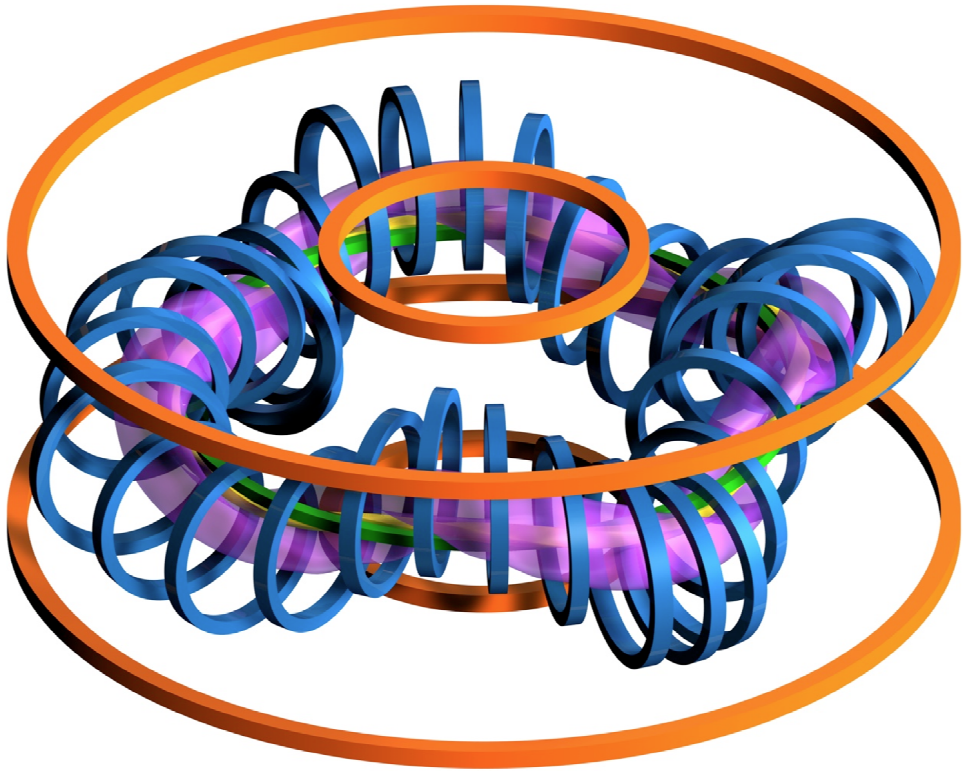

Fig.1 TJ-II coil system. Orange: Ohmic coils. Blue: Toroidal field coils. Yellow: Helical coil. Green central coil. Vertical and radial coil systems sit on the same frame as the ohmic coils, and are not shown in the picture. The boundary of the plasma loop is shown in pink as it twists arround the central conductor.

The currents in the central and helical coils define the standard TJ-II configuration space in terms of rotational transform in vacuum, as shown in Fig 2. This diagram has been computed by choosing the currents in the vertical and toroidal field to maximize the plasma volume while keeping $|\vec{B}|$ constant at the magnetic axis, in order to fulfil the ECRH resonance condition. The TJ-II flexible heliac is able to achieve a wide range of rotational transform values, typically between 0.9 and 2.2, by changing the ratios between the external coil currents, hence the appellative "flexible". The resulting rotational transform has an almost constant radial profile. In addition, transformer induced ohmic current can be used to modify the shear parameter across the plasma, [15], [16], [27],[29],[30],[31], making the TJ-II a device optimally suited for the study of the influence of MHD on confinement and stability in a wide range of rotational transform values. This typically requires numerous plasma discharges with different sets of configuration coil currents being established during the heating/fuelling phase [32],[33], [34], [35], [36]. Plasma discharges are not always easy to reproduce, so the comparison between different types of discharges with different kinetic and magnetic parameters is sometimes cumbersome.

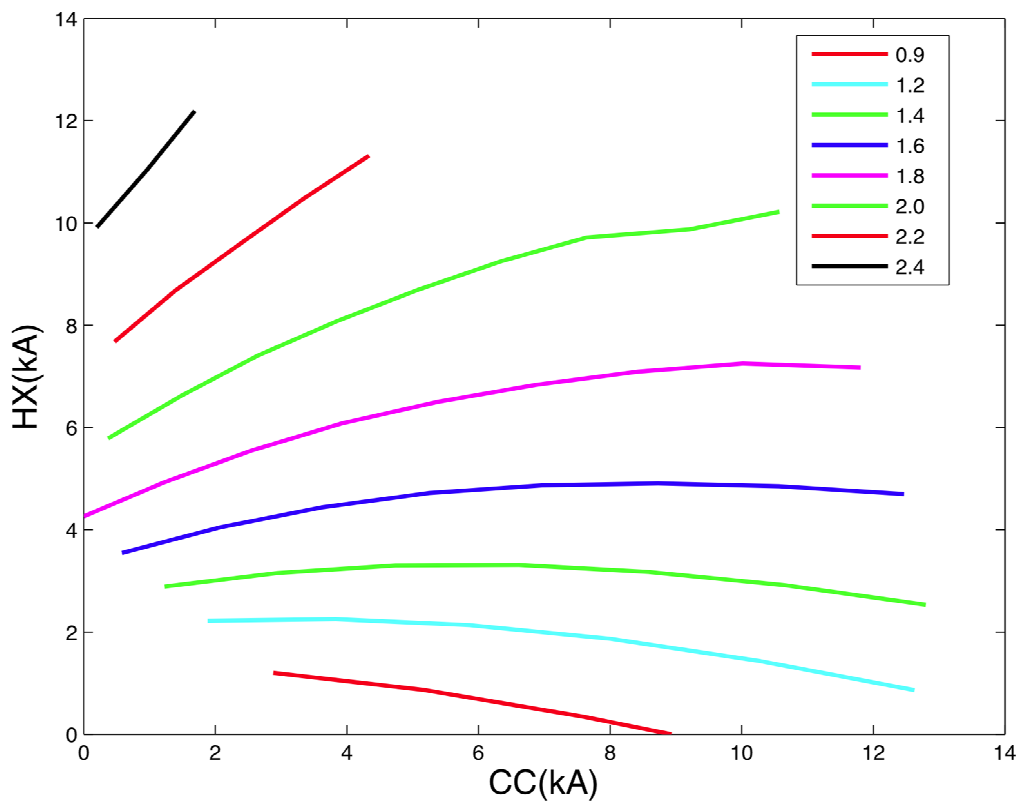

Fig. 2 TJ-II flexibility diagram. Rotational transform contour plots as funtion of the currents in the helical (HX) and central (CC) conductors. Diagram is limited by the maximum currents that can be set in the HX and CC conductors.

The 2008 upgrade of the TJ-II control system [37] [38], [39] allowed the establishment of a rotational transform route in the configuration space that could be swept in a single discharge. Plasma shape, position and rotational transform profiles could then be swept during a single discharge, and equivalence between discharge time and rotational transform could approximately be made. This allowed the comparison between experiments performed on a shot to shot basis [40] with magnetic configuration sweeps [46], for instance. With the exception of the radial field coil, all TJ-II coil systems are inductively coupled to the plasma, so sweeping of the coil configuration currents results in induced ohmic current in the plasma. To avoid this, the transformer coil current was pre-programmed to cancel the total electromagnetic induction at a given radial position inside the plasma, typically the magnetic axis. The idea was that if the electromagnetic induction drive in the plasma could be reduced or eliminated, so it would the plasma current induced during the sweep [39]. The approach was, however, not optimal, since bootstrap current [43],[44], non inductive current drive by the neutral beam injectors, and residual electromagnetic induction arising from imperfect sweep compensation affected the plasma in unpredictable ways, resulting in moderate but uncontrolled excursions of the plasma current respect to the desired reference waveforms. Despite its limitations, the previous control system functionality has been successfully used in MHD studies [46], radial electric field [47], [48] zonal flow and turbulence dynamics [49] [50], supra-thermal ion behaviour [51] etc.

Section II outlines the 2013 upgrade of the TJ-II control system, which has being designed to improve the experiment reproducibility during magnetic configuration sweeps. The control system allows performing feedback on different current signals (see sections III and IV) using the primary transformer coil as the main actuator, while retaining the 2008 upgrade functionality that uses feed-forward waveforms. Vacuum iota profiles can now be held constant or swept in time while maintaining any desired level of toroidal current (within the limitations of the actuator) despite electron temperature perturbations or non-inductive currents present in the plasma. To first order, this new functionality should approximately allow the sweeping of rotational transform and magnetic shear independently, unleashing new experimental possibilities (section V).

## II. CONTROL SYSTEM OVERVIEW

The generic structure of the control system is shown in Fig. 3. Plasma current at TJ-II is measured with a Rogowski type sensor installed inside the stellarator vacuum vessel [54]. In addition, two external Rogowski coils surrounding the vacuum vessel have been installed; a standard design and special winding design to reduce the magnetic interference from surrounding coils [57]. The external Rogowski sensors surround both the vessel and the plasma ring, so according to Ampere's law they measure the sum of vessel and plasma current (torus current).

In order to comply with the high voltage insulation safety requirements, the Rogowski signals are transmitted optically to a front-end computer (the Fast Control System) where are digitally acquired and numerically integrated in real time. To increase the full-scale resolution at the low frequencies of interest, Rogowski signals are low pass filtered on source before being optically transmitted to Fast Control. The vessel current is obtained numerically in Fast Control from the difference between torus current and plasma current.

The signal selected for feedback can be selected from several options; plasma current, torus current, vessel current, and simulated plasma current. The last option is used to ease the commissioning of the digital control system off line, without actually driving real currents through the coils or powering the TJ-II device. All is required from the system is to be able to establish a current level while we perform (or not) a magnetic configuration sweep. All these current feedback options are reviewed in detail in section IV.

The Fast Control system runs the control algorithms and high-level protections with a 1ms sampling rate. This includes a fully programmable Proportional Integral Derivative (PID) digital controller with anti-windup, protections on coil current limits (up to 7.2kA in OH) and ramp rates (less than 50kA/s), as well as failsafe coil current control actions. These are implemented as part of a larger architecture [37] (not detailed in the figure) that initiates the pulse state machine, implements protections in all the coil currents, interfaces with the electrical system issuing real time demands for all the coil currents and, after this new development, controls the plasma, torus or vessel current. Control and safety parameters are fully programmable through web interfaces, and uploaded before the pulse. The OH coil current demand is built from two components; a feed forward component and a control action resulting from the feedback control. All coil current demands are then sent optically to a back-end computer located in the rectifier's control room, ensuring high voltage isolation and noise immunity. The back end computer controls the power supplies directly, as described in [37].

The plasma equivalent circuit is magnetically coupled to the OH transformer and the magnetic configuration coils. A non-inductive current source is shown as part of the plasma circuit to account for one of the main perturbation sources affecting the feedback loop, arising mainly from the bootstrap effect mentioned earlier in the introduction. There is also the possibility to add a fraction of non-inductive current in the plasma using ECCD, but this is not integrated yet as part of the control loop.
The waveforms for plasma and coil currents are pre-programmed according to the magnetic configuration sweep requirements, aiming for approximately independent control of rotational transform and shear. The feed forward waveform for the OH coil current is used to provide with the initial value from which coil current will depart as the result of the additive control action. It can also be used to help the control when fast magnetic configuration sweep experiments are sought after. In this last case, the feed-forward waveform for the OH coil is pre-programmed to approximately cancel out the electromagnetic induction arising from the transients in the configuration coil currents (feed forward sweep compensation). Then, the feedback loop contribution to the OH coil current will add up to the feed forward OH coil current component in order to compensate the residual electromagnetic induction arising from the (always imperfect) feed forward sweep compensation and to approach the desired level of plasma current. The measured current will then try to follow (within the actuator limits) the prescribed current waveform, despite any residual loop voltage or non-inductive current components being present in the plasma.

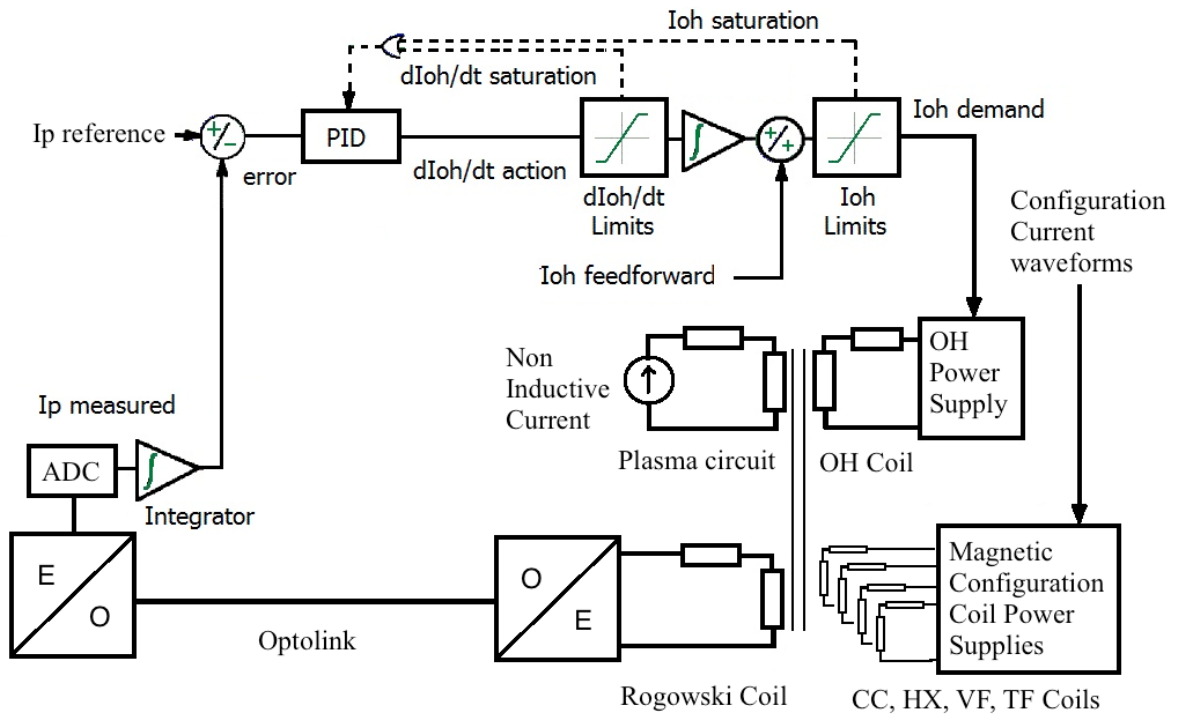

Fig. 3. Control system overview. The signal selected for feedback can be selected from several options; plasma current, torus current, vessel current, and simulated plasma current (only plasma current option is shown in the diagram). The control system is prepared to establish a reference for plasma, torus or vessel current while the coil configuration currents are swept during a single plasma discharge. Only the plasma current feedback option is shown for simplicity. It can also be run in a special simulation mode (which does not require powering the TJ-II device) intended for commissioning and experiment preparation purposes. See main text for details.

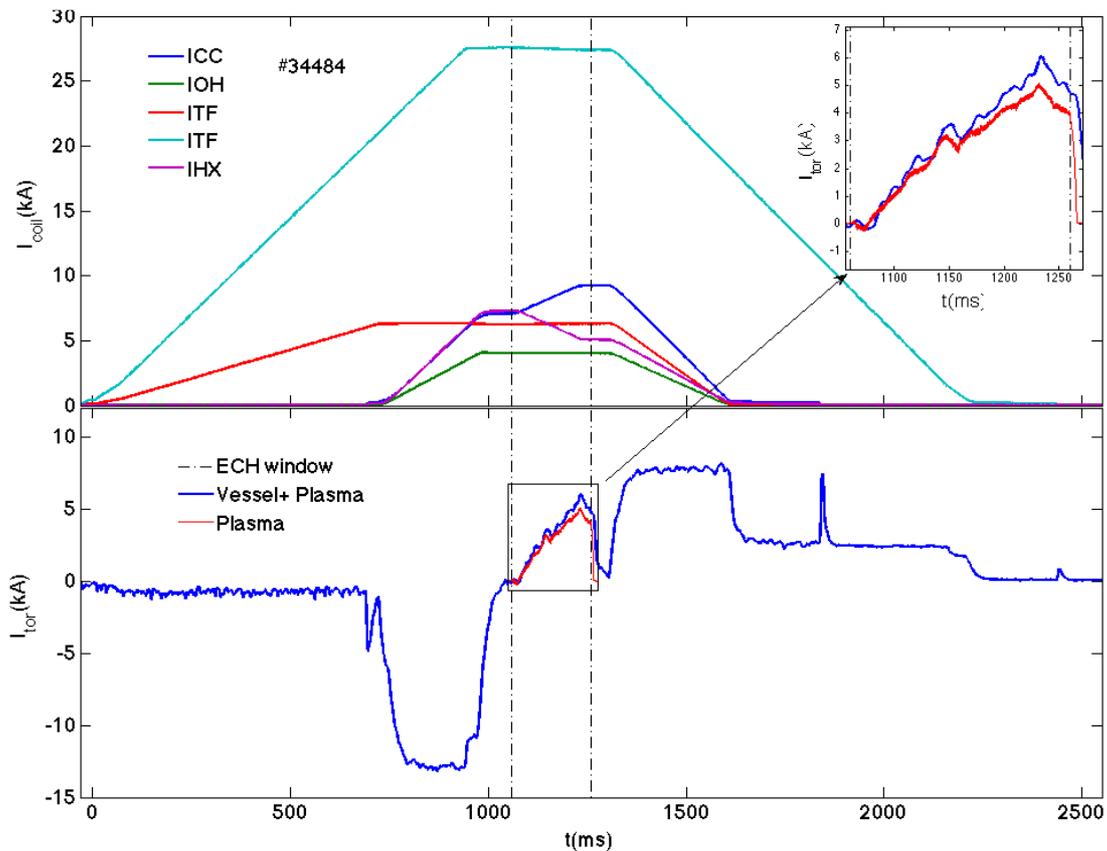

Fig. 4. Top: Configuration and transformer coil currents during a magnetic configuration sweep without compensation. Bottom: Induced current resulting from the sweep as measured by i) a Rogowski sensor encircling both the vacuum chamber and the plasma (blue traces) and ii) a Rogowski sensor placed inside the vacuum vessel and encircling only the plasma ring. The difference between the external and internal measurements is the vessel current, which can account for about 20% of the total plasma current.

# III. CURRENT INDUCTION DURING A MAGNETIC CONFIGURATION SWEEP

To give an idea of typical vessel and plasma current values induced during a magnetic configuration sweep, a TJ-II plasma discharge with induced plasma current is shown in Fig. 4. The discharge has three phases; the ramp-up, ramp-down and flat top. All through this paper the flat-top corresponds with the time window where plasma heating is applied, bearing in mind that in the TJ-II device the magnetic configuration can be swept in time, so the coil configuration currents are not necessarily all "flat" at the "top" of the discharge. This naming convention has been kept in some places, as it is a standard in magnetic fusion research. All through this paper the time window where the Electron Cyclotron Heating is applied will be clearly marked in the figures with vertical discontinuous lines.

During the ramp-up phase, the magnetic configuration is built up gradually, starting with currents in the TF and VF coils, and followed shortly by the CC and HX coil currents. The change in magnetic flux in the ramp-up induces about 13kA of current in the vessel, which decay resistively when the coil configuration currents are stabilized at the beginning of the flat-top. The plasma is formed during the flat-top of discharge, when the ECH is applied. The magnetic configuration sweep takes place shortly after the initial current stabilization. The currents in the central (CC) and helical (HX) coils exhibit the largest changes, since CC and HX dominate the magnetic configuration space (See fig 2). Central and helical coils are also responsible for the largest electromagnetic induction during the magnetic configuration sweeps. For this particular discharge there are about 5kA of plasma current and 6kA of torus current during the sweep, leaving about 1kA of current circulating through the vessel. This is roughly 20% of the total plasma current. The magnetic shear (as determined by plasma current) and rotational transform (as determined by the coil configuration currents) increase roughly linearly with time. The persistent increase plasma current during the sweep links the rotational transform with the magnetic shear, which is a feature that may be desired in some particular type of experiment, but certainly not in general. A more flexible mode of operation that allows decoupling the rotational transform sweep from magnetic shear changes can be obtained by adjusting the transformer coil (OH) current during the sweep. This is achieved by imposing any desired behaviour on plasma current as the magnetic configuration sweep progresses, which is the main subject of this work, and is presented in a later section.

## IV. FEEDBACK OPTIONS FOR MAGNETIC CONFIGURATION SWEEP CONTROL

As we mentioned in section II, the control system allows performing feedback on different current signals (coming from plasma, torus or vessel) despite inductive loop voltage perturbations arising from magnetic configuration sweeps and/or non-inductive current components being present in the plasma. It also allows a special simulation mode for commissioning of future control system upgrades and/or control experiments preparation. These options are summarized in this section.

Using the plasma current feedback option typically involves requesting a null plasma current set point during the sweep. This does not guarantee that the induced ohmic current will be zero in all plasma regions. It only guarantees that the total ohmic current is equal in magnitude and opposite sign to the total bootstrap current. More generally, total plasma current can be forced to follow a prescribed waveform, not necessarily a null set point. Since ohmic current density tends to be distributed in the central plasma regions (where plasma temperature and conductivity is higher), positive ohmic current will generally increase the rotational transform in the central regions of the plasma, while negative ohmic current will do the opposite. The amount of induced plasma current doesn't modify significantly the rotational transform at the last flux surface, even when several kA of plasma current are being induced, so plasma current induction at TJ-II can be assimilated to a method to control the magnetic shear. To do this , an approximate mapping between the desired sweep variables (rotational transform, shear) and the control system references (coils and plasma currents) has to be made. This mapping can be obtained from the rotational transform profiles in vacuum with an added perturbation obtained from a current diffusion simulation using the expected radial profiles for electron temperature, trapped particle fractions and bootstrap current. Several experiments with different levels of plasma current (which will translate in different levels of shear) and configuration coil currents (which will translate in different levels of rotational transform) can then be designed to scan for different shear-dependent and/or rotational transform dependent plasma effects, but with the added advantage of good experimental reproducibility arising from the feedback control.

The scheme is, however, far from perfect. There is no rotational transform diagnostic at TJ-II, so the resulting rotational transform obtained in this type of experiments is still uncertain. At present, the rotational transform profiles of a given experiment have to be inferred from the theoretical expectations (e.g. vacuum rotational profiles, current diffusion) and MHD activity as measured with magnetic and radiation diagnostics [58] . A new Motional Stark Effect (MSE) diagnostic is being developed at TJ-II, which will help to infer the rotational transform more accurately in the future [59].

Another variable made available for feedback control is the torus current. When torus current is zero in steady state, the sum of plasma current and vessel current must be of opposite sign to the non-inductive current component (e.g. bootstrap). As with plasma current feedback control, this does not guarantee that the induced ohmic current will be zero in all plasma regions, but just as before, the torus current can be forced to follow a prescribed waveform, which will modify the magnetic shear during the experiments in, admittedly uncertain, but certainly more reproducible ways. The next section will show the experimental results using this option.

There is another interesting possibility, which is to feedback directly on a vessel current signal. The field created by the vessel currents is in general, non-stellarator symmetric, so it could in principle perturb the vacuum configuration in undesirable ways. By forcing the vessel current to zero during the sweep we can make sure that the vessel current does not modify the nominal vacuum magnetic configuration. This does not impose any restriction on the plasma current that is free to float during the sweep. This option has been included to improve de reproducibility of the ECH plasma breakdown right at the beginning of the flattop, but has not been tested at the time of writing this paper.

Finally, the control system can be run in simulation mode. Using this option, the torus current is simulated using a 2$^{nd}$ order discrete state space model of the form:

$$\begin{aligned}\mathbf{x}(t+\Delta t) &= \mathbf{A}\cdot\mathbf{x}(t)+\mathbf{B}\cdot\mathbf{u}(t)\\ y(t) &= \mathbf{C}\cdot\mathbf{x}(t)+\mathbf{D}\cdot\mathbf{u}(t)\end{aligned} \qquad (2)$$

Where $\mathbf{x}(t)$ is a four dimensional state space vector, y(t) is the torus current and $\mathbf{u}(t)$ is a vector containing the currents in all the coil systems but the radial field coil system (which does not contribute to the induction process). The numerical values for the matrices $\mathbf{A}$, $\mathbf{B}$, $\mathbf{C}$, $\mathbf{D}$ are obtained from actual experiments using system identification techniques [60]. To this extent, a database of 36 plasma discharges has been used.

While in simulation mode, feedback is performed on the simulated torus current signal. The resulting control action for the OH coil is used recursively as inputs to the

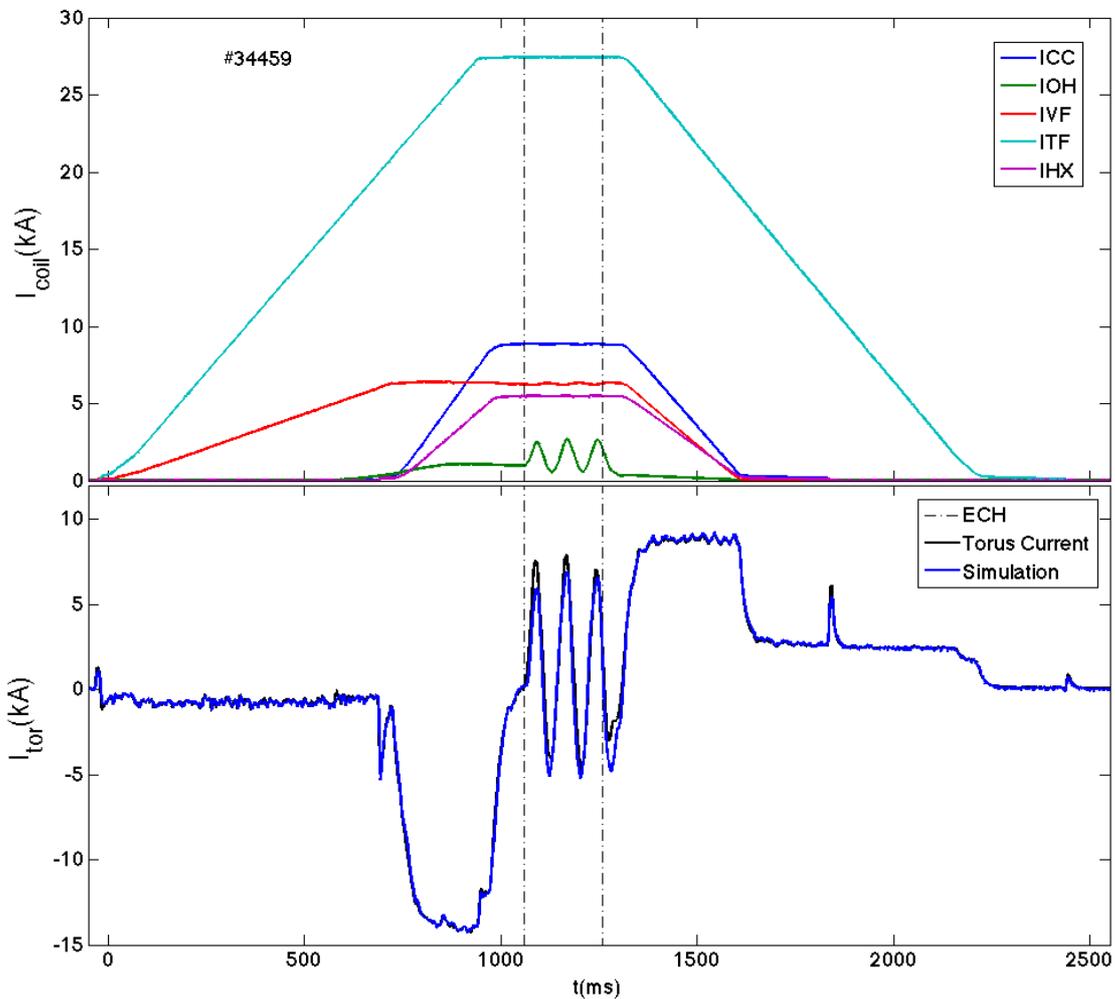

Fig. 5 Torus current simulation using a 2nd order state space model. Inputs to the model are the coil currents, shown in the top figure. The simulations results are shown in blue in the bottom figure, along with the torus current measurement (black) for comparison.

simulation model, along with all the pre-programmed coil current references. This mode of operation has been introduced mainly for commissioning purposes, as it does not require powering the TJ-II device. It can also be used to compute approximations to the OH feed forward waveforms when complex magnetic configuration sweep experiments

(e.g. involving fast transients) are being prepared. Figure 5 shows the simulation results using this model for an actual plasma discharge. This level of agreement is extensible to all the 36 discharges checked in the database; this gives enough confidence to serve its intended purpose for commissioning and experiment preparation.

## V. MAGNETIC CONFIGURATION SWEEP CONTROL EXPERIMENTS

From the experimental point of view, both plasma and torus current can be used indistinctly for control purposes, since all is required from the system is to be able to establish a current level while we perform (or not) a magnetic configuration sweep. The reference current will be somehow related to a shear modification in the plasma, as discussed earlier in the previous section.

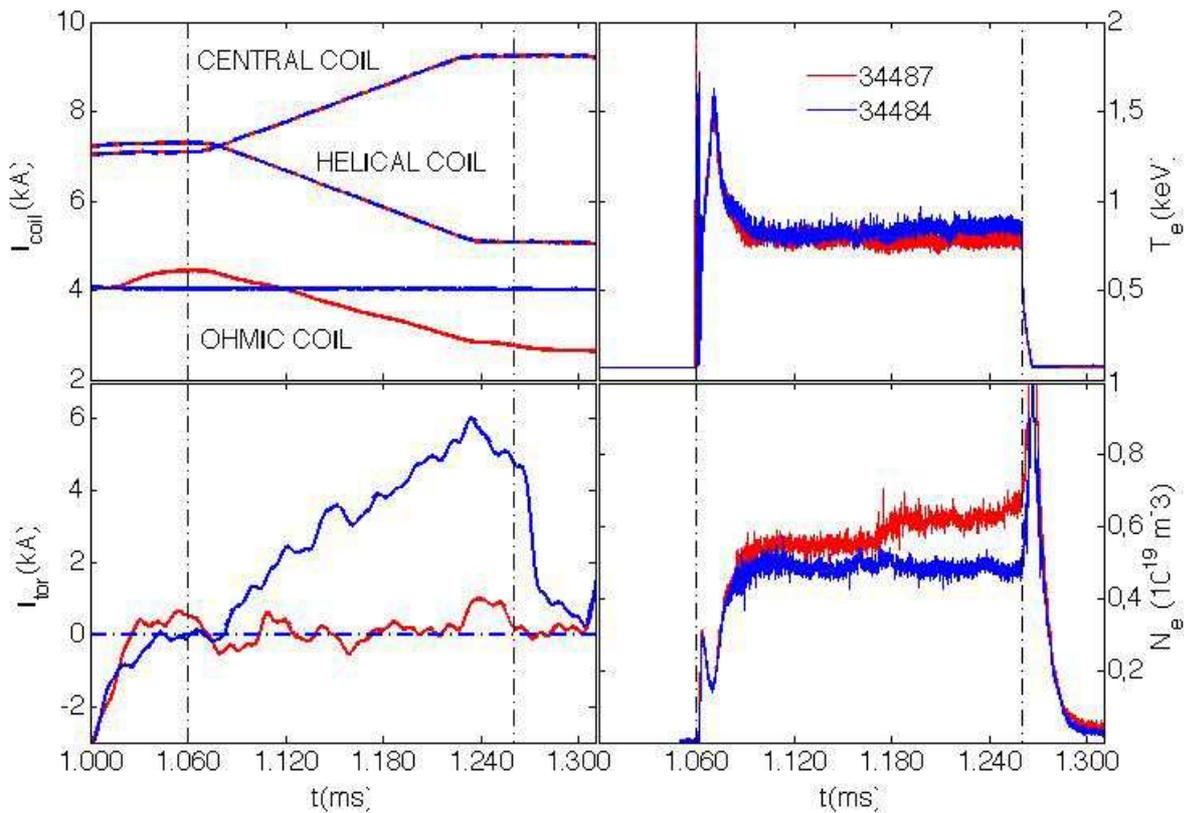

Fig. 6 It is shown how torus current can be kept close to zero during a magnetic configuration sweep, despite plasma discharge evolution in terms of temperature and line-averaged density. Top left: Currents in the helical and central conductors during a magnetic configuration sweep without torus current control (blue traces) and with torus current control (red traces). Bottom left: Resulting torus current induced during the sweep for both cases. Top right: Central electron temperature as measured by electron cyclotron emission. Bottom right: Line averaged density as measured by a microwave interferometer.

From point of view of experiment preparation, however, plasma current signal is generally preferable. The torus current signal contains a vessel current component that can be as high as 20% of the total plasma current, so mapping between the desired shear and torus current is more difficult than the mapping between shear and plasma current. Plasma current would be obviously our first choice. However, the plasma current signal could not be used reliably when the control system was being commissioned, due to some capacitive coupling between the Rogowski leads and the voltage applied to the TF coil.   Because of this, all the experiments presented in this section use the torus current instead.

Fig. 6 shows how the torus current feedback loop can be used to compensate the electromagnetic induction in a magnetic configuration sweep, our reference sweep from now on. For all the windows shown the vertical discontinuous lines mark the time window where the Electron Cyclotron Heating is applied and the plasma is formed. Before the ECH window, the torus current contains exclusively the vessel current component, since there is no plasma in this phase. A resistive current decay can be appreciated in the torus current signal a few milliseconds before the feedback is applied.

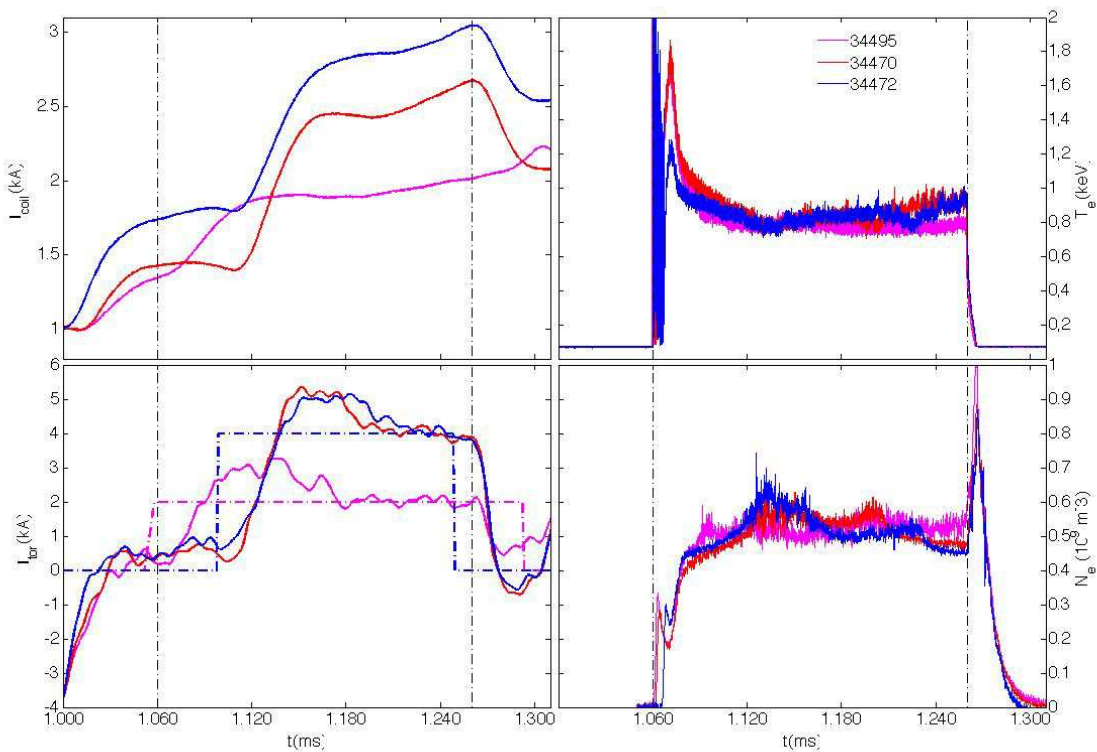

Fig. 7 Torus current step responses in standard flat top discharges without magnetic configuration sweep.  Using identical torus current references (bottom left), the control system is acting differently on the OH coil current (top left) in order to compensate for slightly different plasma temperature (top right) and density (bottom right) evolution.

This is part of the typical signature of a discharge ramp-up (see for instance Figs. 4, 5). The currents in the helical, central and transformer coils during a magnetic configuration sweep are shown in the top left window. Without any kind of compensation (blue traces), there are about 6kA of torus current induced during the sweep.  Using the feedback loop (red traces) the torus current (bottom left) is kept at the null set point. This is performed despite plasma discharge evolution in terms of temperature (top right) and line averaged density (bottom right).

Fig. 7 shows positive current step responses for standard flat-top discharges without magnetic configuration sweeps. Two identical +4kA torus current step responses (bottom left window) are shown to illustrate how the control system is acting differently in the OH coil current (top left) in order to compensate for slightly different plasma temperature (top right) and density (bottom right) evolution.

Fig. 8 shows null torus current regulation (blue traces) and -4kA step response (red traces) for a standard discharge without magnetic configuration sweep. A negative (-2kA) step response (pink traces) performed during the reference sweep is also shown to illustrate the capability of the system to establish magnetic configuration sweep and torus current independently, a major feature of the control system. It can also be observed how the control on the torus current can be achieved despite variable electron temperature and density.

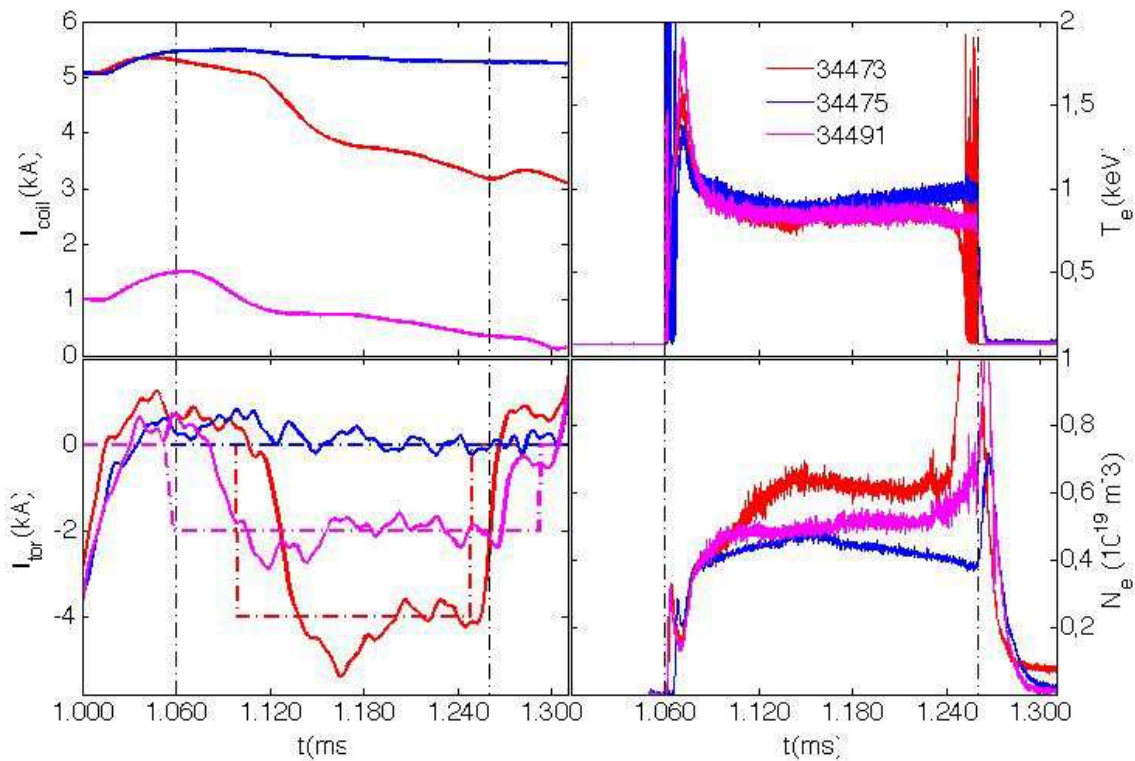

Fig. 8. Torus current step response for several values of negative plasma current. The current in the OH coil is shown in the top left. The resulting torus current and references in the bottom left corner. Plasma temperature (top right) and line averaged density (bottom left) are show to show how the control system is able to cope with variable plasma conditions. The null current and -4kA step reference were run in standard flat-top discharges without magnetic configuration sweep. The step response corresponding to -2kA was run during the magnetic configuration sweep of figure 6, to illustrate the capability of the system to establish magnetic configuration sweep and torus current independently.

# VI. CONCLUSIONS

A magnetic configuration sweep control system has been developed for the TJ-II Heliac device. The system is prepared to establish a reference for plasma, torus or vessel current while we perform (or not) a magnetic configuration sweep. It can also be run in a special simulation mode that does not require powering the TJ-II device, intended for commissioning and experiment preparation purposes. Preliminary tests with the control system have shown its ability to perform magnetic configuration sweeps while establishing a torus current independently, despite variable plasma conditions. This result opens new experimental possibilities to scan the rotational transform (as determined from coil configuration currents) independently of the shear (as determined by plasma current).

# VII. ACKNOWLEDGMENTS

The Spanish Ministry of Science and Innovation (MICINN) has supported this work through Research Project ENE2010-18345.